\documentclass[3p,times,twocolumn]{elsarticle}

\usepackage{ecrc}

\volume{00}

\firstpage{1}

\journalname{Nuclear Physics B Proceedings Supplement}

\runauth{}

\jid{nuphbp}

\jnltitlelogo{Nuclear Physics B Proceedings Supplement}

\usepackage{amssymb}

\usepackage[figuresright]{rotating}
\usepackage{multirow}
\usepackage{caption}
\usepackage{subcaption}
\usepackage{lineno} 

\begin{document}
\begin{frontmatter}



\dochead{}

\newcommand{\ttbar}{\ensuremath{t\bar{t}}}
\newcommand{\tantit}{top-antitop quark}
\newcommand{\tquark}{top quark}
\newcommand{\tquarka}{top-quark}
\newcommand{\xsec}{cross section}
\newcommand{\xseca}{cross-section}

\title{Top-quark mass measurements using the ATLAS detector at the LHC}


\author{Kaven Yau Wong, Physikalisches Institut, Universit\"at Bonn, On behalf of the ATLAS Collaboration}

\address{}

\begin{abstract}
The most recent results of the \tquarka\ mass measurements with the ATLAS detector using data collected from 
proton-proton collisions at the Large Hadron Collider are presented. 
Although several decay modes of the \tquarka\ pairs have been used in ATLAS for \tquarka\ mass measurements,
only the latest results are presented (single lepton and dilepton channels).
The \tquarka\ pole mass from the \ttbar\ \xseca\ measurement in the dilepton channel and 
the \tantit\ mass difference measurement in the single-lepton channel are also shown.
The systematic uncertainties associated to these measurements are discussed in some detail.
\end{abstract}


\end{frontmatter}


\newcommand{\ATLASC}{ATLAS Collaboration}


\newcommand{\tquark}{top quark}
\newcommand{\Tquark}{Top quark}
\newcommand{\tquarka}{top-quark}
\newcommand{\WBoson}{\ensuremath{W} boson}
\newcommand{\WBosona}{\ensuremath{W}-boson}
\newcommand{\tantit}{top-\antitop}
\newcommand{\Tantit}{Top-\antitop}
\newcommand{\antitop}{antitop}
\newcommand{\antitquark}{\antitop\ quark}
\newcommand{\antitquarka}{\antitop-quark}

\newcommand{\Tquarka}{Top-quark}
\newcommand{\SM}{Standard Model}
\newcommand{\ATLASCol}{ATLAS Collaboration}
\newcommand{\MC}{Monte-Carlo}
\newcommand{\ie}{\textit{i.e.}}
\newcommand{\xsec}{cross section}
\newcommand{\xseca}{cross-section}
\newcommand{\cofM}{centre-of-mass}

\newcommand{\bjet}{\ensuremath{b}-jet}
\newcommand{\btag}{\ensuremath{b}-tag}
\newcommand{\btagged}{\btag ged}
\newcommand{\btagging}{\btag ging}
\newcommand{\bJSF}{\ensuremath{b}JSF}

\newcommand{\refFig}[1]{Figure \ref{#1}}
\newcommand{\refTable}[1]{Table \ref{#1}}
\newcommand{\sqrts}[2]{\ensuremath{\sqrt{s}=#1\ #2}}

\newcommand{\kFigWidth}{0.48}
\newcommand{\kMiniPageWidth}{0.48}

\def\TeV{\ifmmode {\mathrm{\ Te\kern -0.1em V}}\else
                   \textrm{Te\kern -0.1em V}\fi}%
\def\GeV{\ifmmode {\mathrm{\ Ge\kern -0.1em V}}\else
                   \textrm{Ge\kern -0.1em V}\fi}%

\def\ifb{\mbox{fb\ensuremath{^{-1}}}}

\newcommand{\ttbar}{\ensuremath{t\bar{t}}}
\newcommand{\Rlb}{\ensuremath{R_{\ell b}^{\mathrm{reco}}}}
\newcommand{\mass}[2]{\ensuremath{m_{#1}^{\mathrm{#2}}}}
\newcommand{\mtop}[1]{\mass{\mathrm{top}}{#1}}
\newcommand{\mlb}{\mass{\ell b}{}}
\newcommand{\mDiffFit}{\ensuremath{\Delta_m^\mathrm{fit}}}

\newcommand{\titleSingleLepton}{\Tquarka\ mass measurement in the \ttbar\ single lepton channel}
\newcommand{\titleDilepton}{\Tquarka\ mass measurement in the \ttbar\ dilepton channel}
\newcommand{\titlePoleMass}{\Tquarka\ pole mass from the \ttbar\ \xseca\ measurement in the dilepton channel}
\newcommand{\titleMassDiff}{\Tantit\ mass difference in the \ttbar\ single lepton channel}

\section{Introduction}
The \tquark\ is the heaviest elementary particle in the \SM\ and its mass is a fundamental parameter in quantum chromodynamics.
Its value must be determined experimentally and its precise measurement has a large impact in the computation of electroweak corrections.

The ATLAS detector \cite{ATLAS} is a general purpose detector located at the Large Hadron Collider (LHC) \cite{LHC}.
Over the last years, the \ATLASCol\ has measured the \tquarka\ mass with increasing precision,
with the following most recent analyses:
\begin{itemize}
  \item \titleSingleLepton\ \cite{SingleLepton};
	\item \titleDilepton\ \cite{Dilepton};
	\item \titlePoleMass\ \cite{PoleMass};
	\item \titleMassDiff\ \cite{MassDiff}.
\end{itemize}

Almost all the analyses use the template method to extract the variable of interest from data,
where the templates are built with the help of \MC\ simulations.

\section{\titleSingleLepton}
The analysis \cite{SingleLepton} is performed using data at a \cofM\ energy of \sqrts{7}{\TeV},
which amounts to an integrated luminosity of 4.7 \ifb.

A three-dimensional template technique is used, 
where the \mtop{reco}, the jet scale factor (JSF) and the \bjet\ scale factor (\bJSF) are measured simultaneously
by fitting the distribution of three observables reconstructed using a kinematic likelihood fit: 
the \tquarka\ mass, the \WBosona\ mass and 
the ratio between the average transverse momentum of the \btagged\ jets 
and the average transverse momentum of the two jets of the \WBosona\ hadronic decay, \Rlb (see \refFig{Figs1L}).

The reconstructed \tquarka\ mass is expected to be sensitive to the \tquarka\ mass, the JSF and the \bJSF,
while the reconstructed \WBosona\ mass is expected to only depend on the JSF.
Finally, the reconstructed $R_{lb}$ is expected to depend on the \bJSF\ and the \tquarka\ mass.

\begin{figure}
  \begin{minipage}{\kMiniPageWidth\textwidth}
    \includegraphics[width=0.99\textwidth]{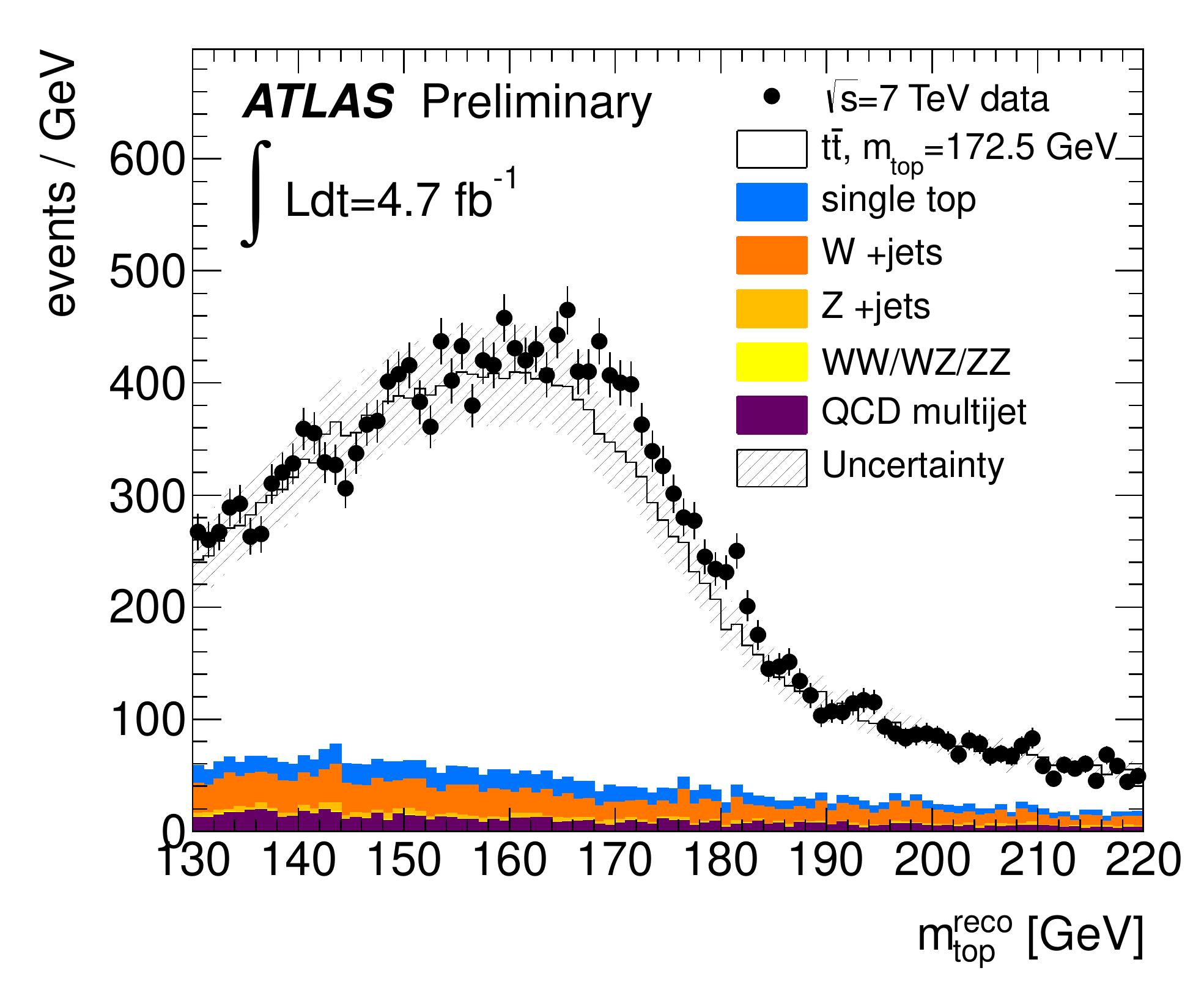}
  \end{minipage}
  \begin{minipage}{\kMiniPageWidth\textwidth}
    \includegraphics[width=0.99\textwidth]{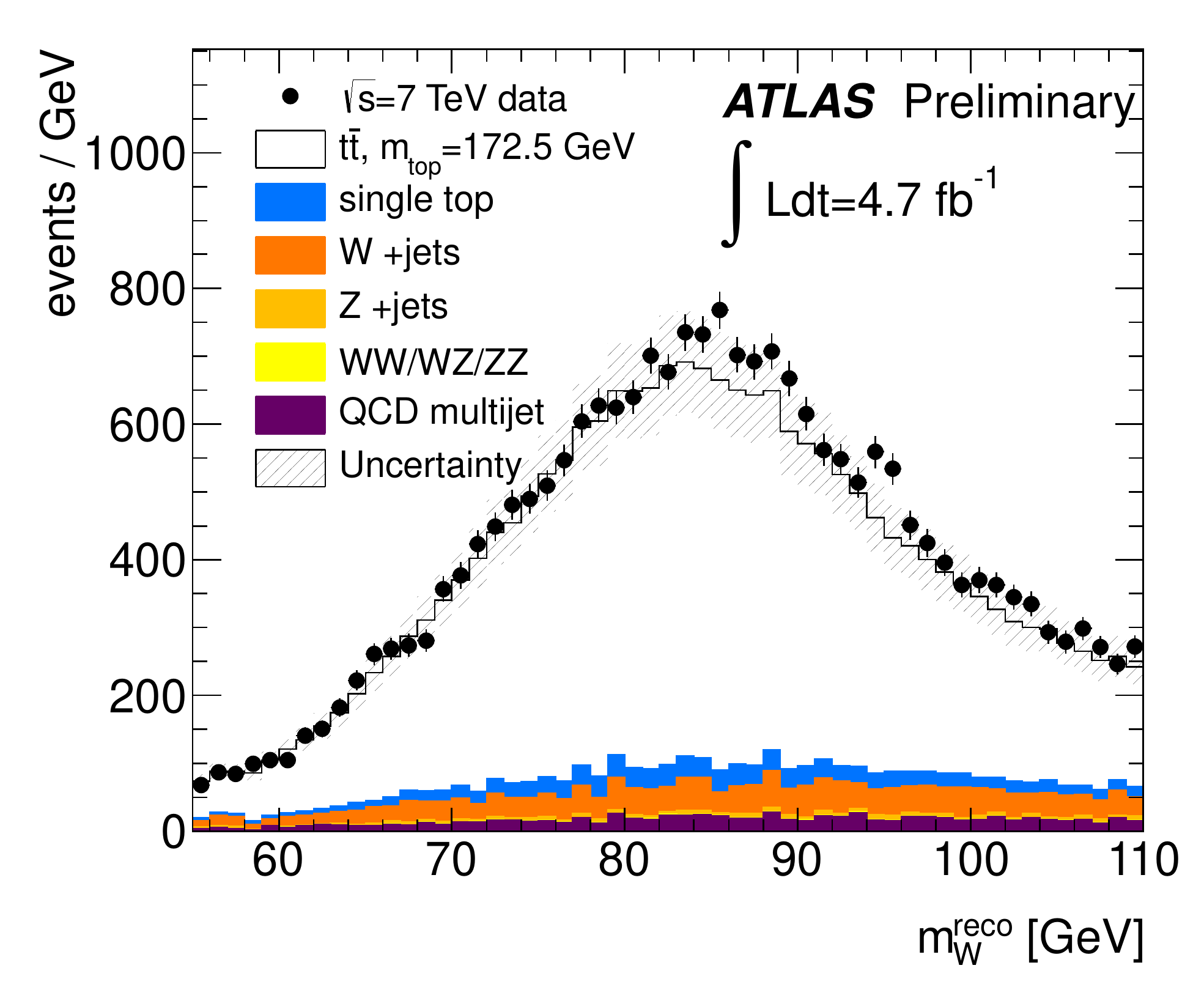}
  \end{minipage}
  \begin{minipage}{\kMiniPageWidth\textwidth}
    \includegraphics[width=0.99\textwidth]{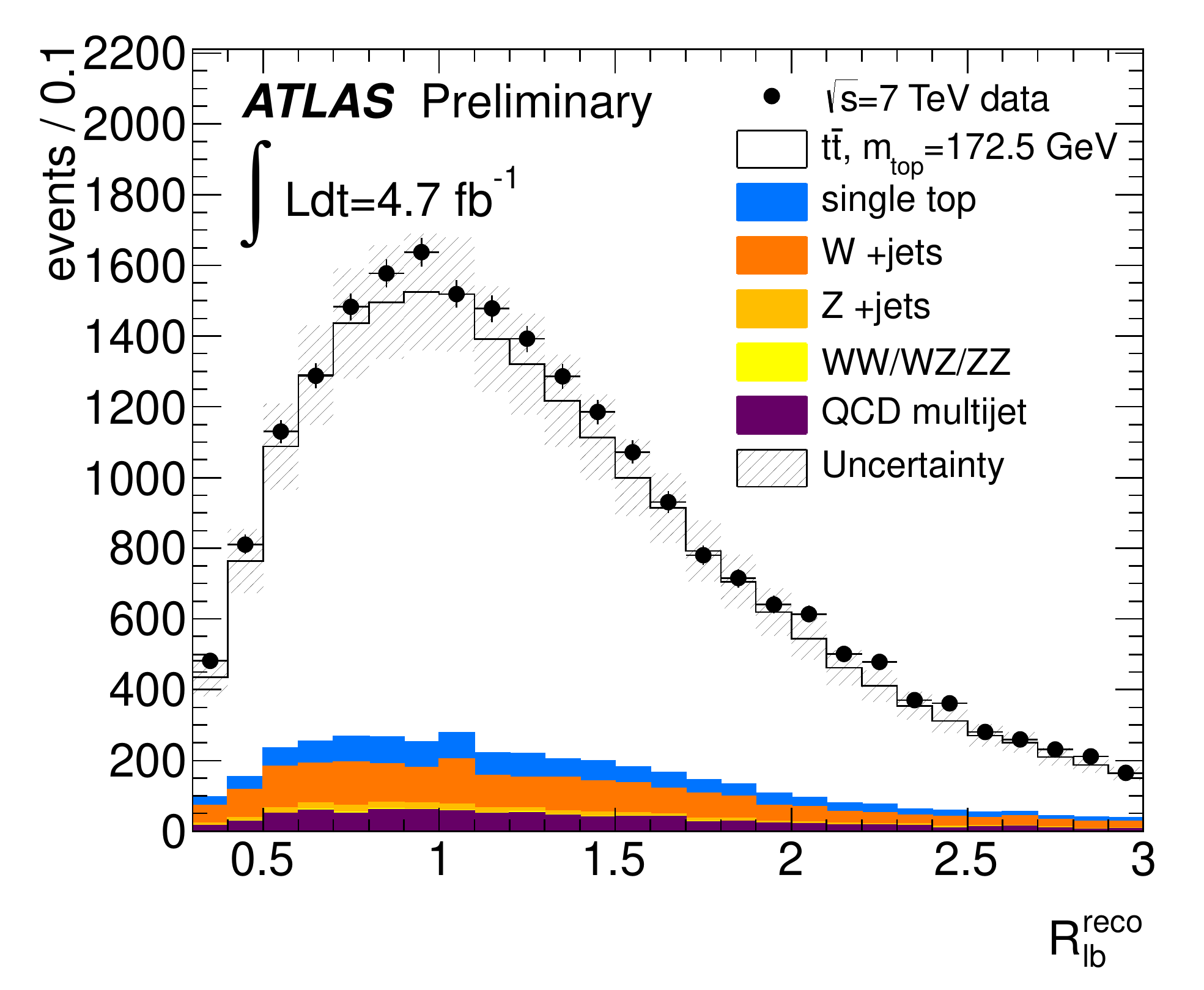}
  \end{minipage}
	\caption{Data-MC comparison for \mtop{reco}\ (top), \mass{W}{reco}\ (middle) and \Rlb\ (bottom).
	Each point is obtained from the result of the per-event kinematic likelihood fit.
	The hashed area is the total uncertainty.
	}
	\label{Figs1L}
\end{figure}

Applying the three-dimensional template technique to data, the \tquarka\ mass is measured to be: 
\begin{equation}
  \mtop{}= 172.31 \pm 0.75 \mathrm{(stat)} \pm 1.35 \mathrm{(syst)} \GeV,
\end{equation}
where the statistical uncertainty also include the uncertainty from the JSF and \bJSF\ measurements.

By comparing a two-dimensional template analysis (where the \bjet\ energy scale is fixed) to 
the three-dimensional template analysis (where the \bjet\ energy scale is allowed to vary),
it is shown that the three-dimensional template technique significantly reduces 
the \bjet\ energy scale uncertainty from $0.92 \GeV$ to $0.08 \GeV$ 
and the hadronization uncertainty from $1.30 \GeV$ to $0.27 \GeV$.  
There is a drawback, however, since the \btagging\ efficiency and mistag rate uncertainty 
increases from $0.17 \GeV$ to $0.81 \GeV$.
But the overall effect is a total systematic uncertainty improvement of 33\%, reducing from $2.02 \GeV$ to the final $1.35 \GeV$ quoted in this measurement.

\section{\titleDilepton}
The dilepton channel does not allow a direct mass reconstruction, 
but has the advantage that it offers a very clean signal and a very good signal-to-background ratio.
The analysis \cite{Dilepton} is performed using 4.7~\ifb\ of 7~\TeV\ data.

The template method is used to measure the \tquarka\ mass using the \mlb\ variable, 
defined as the average invariant mass of the two lepton-\bjet\ systems in the dileptonic channel.
Since the correct pairing between the two leptons and the two \bjet s is not known,
the values of \mlb\ for both combinations are computed and the smallest value is taken.
This algorithm gives the correct pairing 77\% of the times.

The \mlb\ distribution of the signal is modelled as the sum of a Gaussian function and a Landau function for the signal, while th background is modelled as a Landau function.
It is to be noted that the background is very small, contributing only 3\% to the total number of events.  
The template fit function for different input \tquarka\ masses is shown in \refFig{2LTemplate}.
\begin{figure}[h]
  \includegraphics[width=\kFigWidth\textwidth]{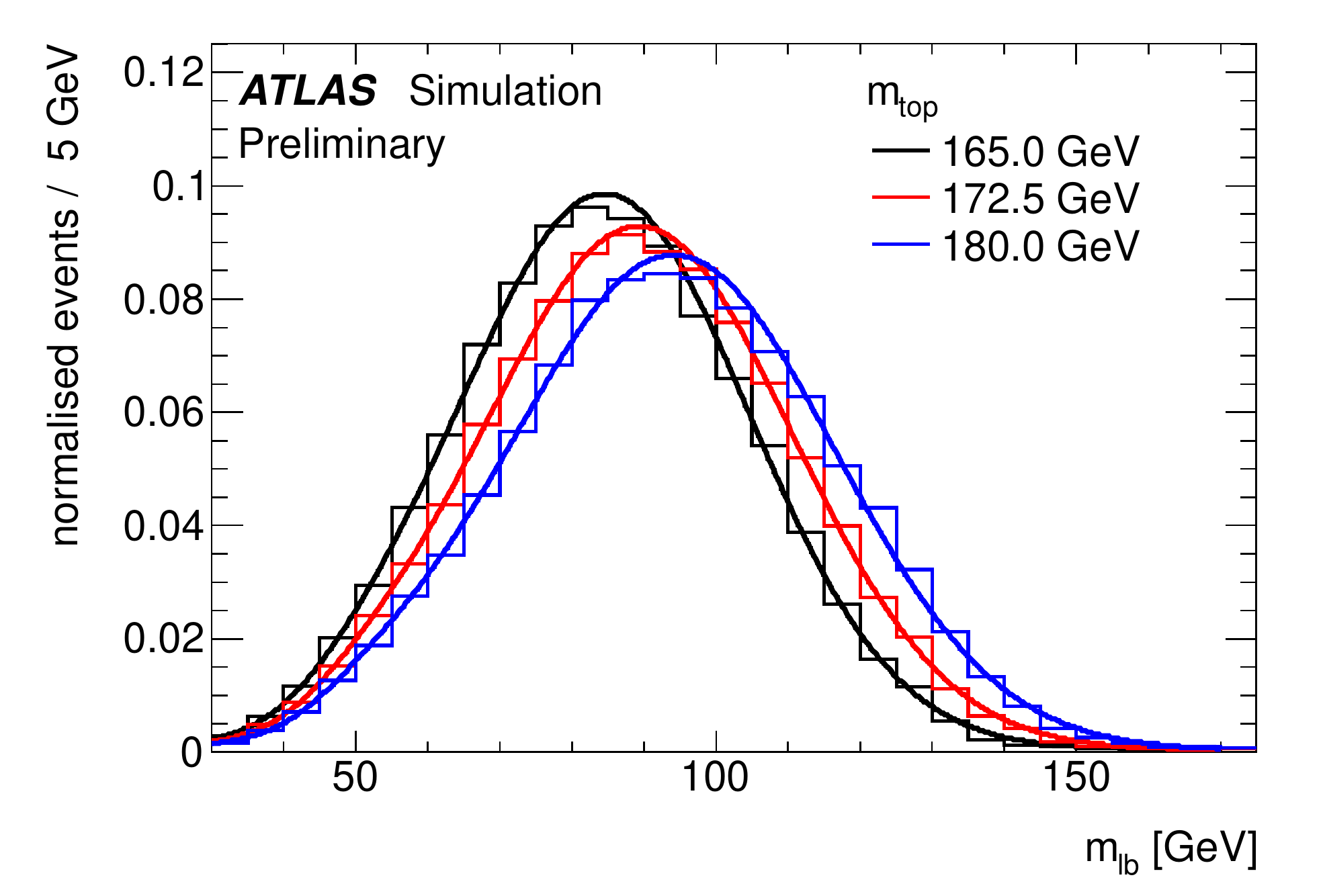}
	\caption{\mlb\ template fit function for different input \tquarka\ masses.  The plot shows the sensitivity of \mlb\ to the \tquarka\ mass.}
	\label{2LTemplate}
\end{figure}

Applying the template fit to data gives the value (see \refFig{2LFitData}): 
\begin{equation}
  \mtop{}= 173.09 \pm 0.64 \mathrm{(stat)} \pm 1.50 \mathrm{(syst)} \GeV,
\end{equation}
where the systematic uncertainty is dominated by the jet energy scale ($0.89 \GeV$) and the \bjet\ energy scale ($0.71 \GeV$).

\begin{figure}[h]
  \includegraphics[width=\kFigWidth\textwidth]{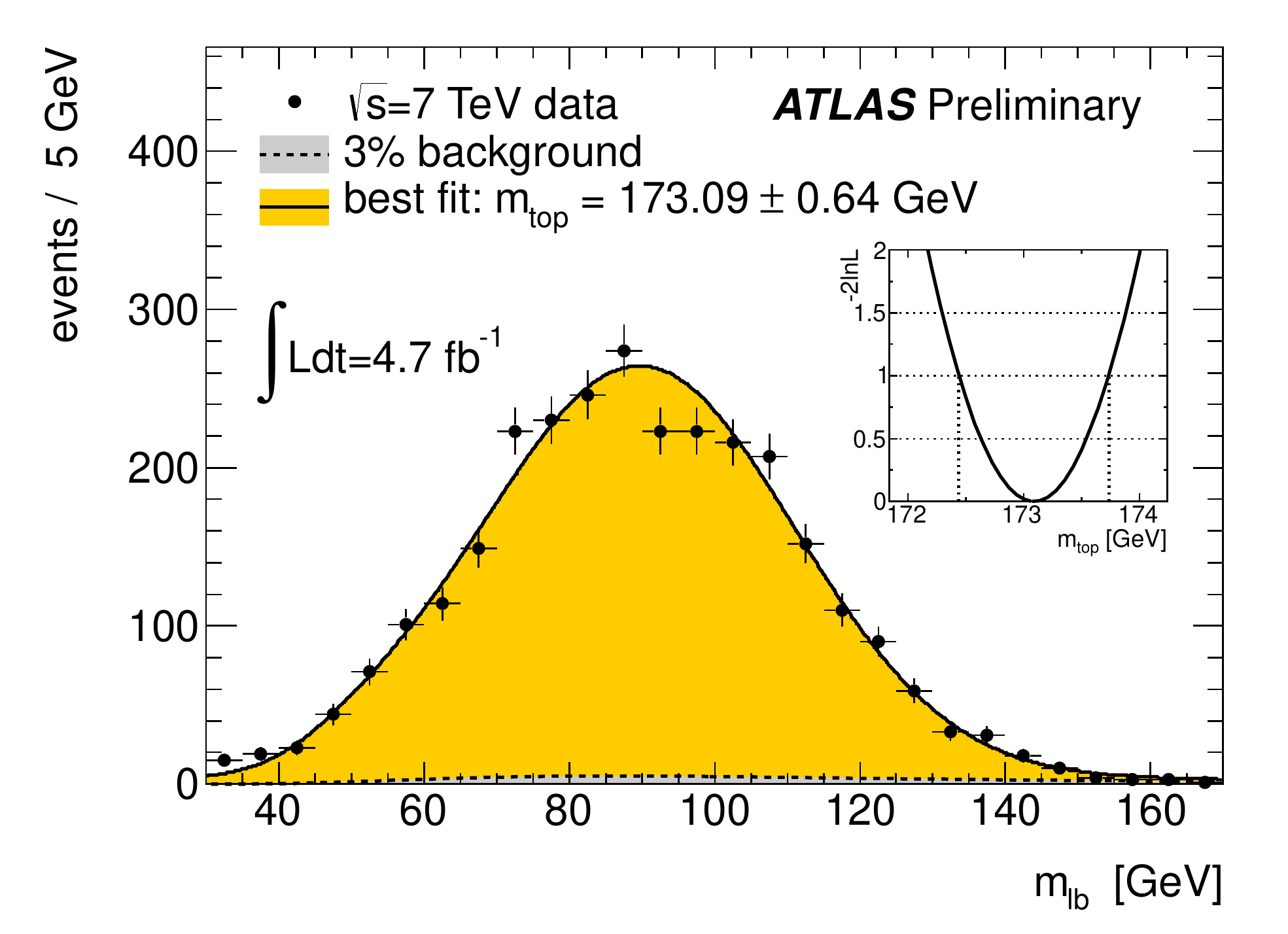}
	\caption{\Tquarka\ mass measured by fitting the template function to data.	Only the statistical uncertainty is quoted.}
	\label{2LFitData}
\end{figure}

\section{\titlePoleMass}
The ATLAS collaboration has performed a \ttbar\ \xseca\ measurement in the dilepton channel using 7~\TeV\ and 8~\TeV\ data,
amounting to integrated luminosities of 4.6~\ifb\ and 20.3~\ifb\ respectively \cite{PoleMass}.

With the \tquarka\ mass measurements reaching precisions of the order $1 \GeV$, 
one should remember that the values of these reconstructed \tquarka\ masses
are different from the value of the \tquarka\ pole mass.
This difference has been estimated to be of the order of $1 \GeV$ \cite{TopMassMCVsPole}.

The \tquarka\ pole mass, \ie, the mass of the \tquark\ as a free particle, can be computed from the \ttbar\ \xsec.
The idea is to exploit the strong dependence of the \ttbar\ \xsec\ on the \tquarka\ pole mass.
This dependence is obtained for both \cofM\ energies using different parton density function (PDF) models, as shown in \refFig{poleTheoDep}.
Notice that the measured \xsec\ does not depend on the assumed input \MC\ \tquarka\ mass,
ensuring that the method used to measure the \ttbar\ \xsec\ is independent from the value of the \tquarka\ pole mass.
This requirement is important in order to extract the \tquarka\ pole mass from the measured value of the \ttbar\ \xsec.

The dependence of the \ttbar\ \xsec\ on the \tquarka\ pole mass is parametrized as:
\begin{equation}
  \ensuremath{\sigma_{\ttbar}^\mathrm{theo} \left(\mtop{pole}\right)= \sigma\left(\mtop{ref}\right)\left(\frac{\mtop{ref}}{\mtop{pole}}\right)^4\left(1+a_1 x+a_2 x^2\right)},
\end{equation}
where $\mtop{ref}=172.5 \GeV$, \ensuremath{x=\left(\mtop{pole}-\mtop{ref}\right)/{\mtop{ref}}}, and \ensuremath{\sigma\left(\mtop{ref}\right)}, \ensuremath{a_1} and \ensuremath{a_2}
being free parameters.

\begin{figure}[h]
  \includegraphics[width=\kFigWidth\textwidth]{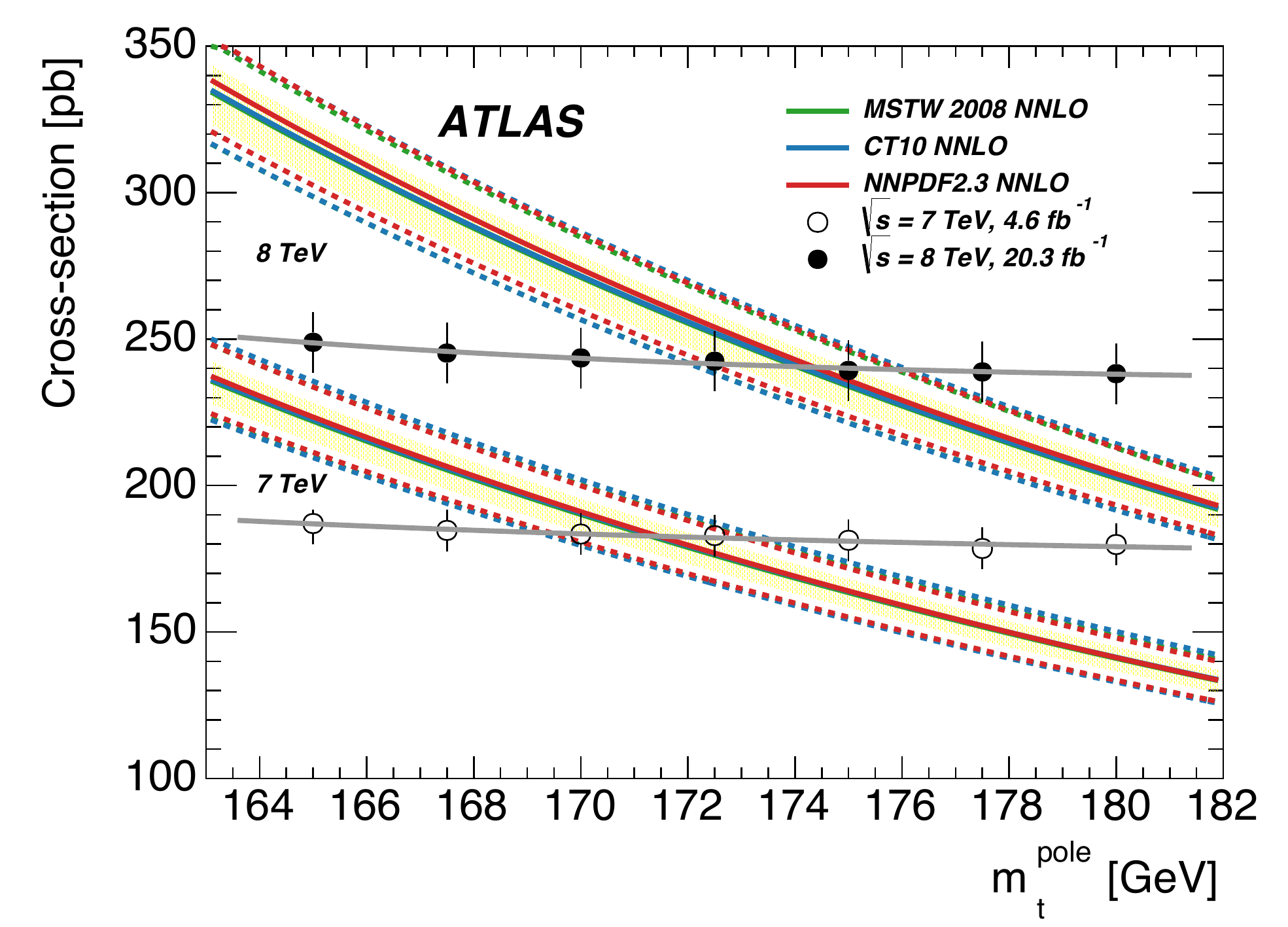}
	\caption{
	Predicted NNLO+NNLL \ttbar\ production cross-sections at \sqrts{7}{\TeV} and \sqrts{8}{\TeV} as a function of \mtop{pole} using different PDF sets.
	The total uncertainty is shown in dashed lines, while the yellow band shows the QCD scale uncertainty.
	}
	\label{poleTheoDep}
\end{figure}

Once the parameters are computed, the pole mass is extracted by maximizing the likelihood:
\begin{equation}
  \ensuremath{
	  \mathcal{L}=\int{
		  G\left(\sigma_{\ttbar}' \vert \sigma_{\ttbar}, \rho_{exp}\right)
		  G\left(\sigma_{\ttbar}' \vert \sigma_{\ttbar}^{theo}, \rho^{\pm}_{theo}\right)
			d{\sigma_{\ttbar}'}
		}
	},
\end{equation}
where \ensuremath{\mathcal{L}}, \ensuremath{\sigma_{\ttbar}} and \ensuremath{\sigma_{\ttbar}^{theo}} depend on \mtop{pole}. 
The function \ensuremath{G\left(x \vert \mu,\rho\right)} 
is a Gaussian probability density in the variable \ensuremath{x} 
with mean \ensuremath{\mu} and standard deviation \ensuremath{\rho}.

\begin{table}[h]
  \begin{tabular}{ccc}\hline
	  \multirow{2}{*}{PDF} & \multicolumn{2}{c}{\mtop{pole} (\GeV) from $\sigma_{\ttbar}$} \\ 
		                     & \sqrts{7}{\TeV}    & \sqrts{8}{\TeV}  \\ \hline
    CT10 NNLO            & $171.5 \pm 2.6$    & $174.2 \pm 2.6$ \\ 
		MSTW 68\% NNLO       & $171.4 \pm 2.4$    & $174.0 \pm 2.5$ \\ 
		NNPDF2.3 5f FFN      & $171.4 \pm 2.3$    & $174.2 \pm 2.4$ \\ \hline
	\end{tabular}
	\caption{\mtop{pole} measurements using different PDF sets.}
	\label{poleMeasurement}
\end{table}

The \tquarka\ pole mass measurements using different PDF sets and the two different \xsec s are shown in \refTable{poleMeasurement}.
By combining both measurements, the \tquarka\ pole mass is found to be: 
\begin{equation}
  \mtop{pole}= 172.9^{+2.5}_{-2.6} \GeV.
\end{equation}

\section{\titleMassDiff}
The analysis \cite{MassDiff} is performed in the \ttbar\ single-lepton channel
using data at a \cofM\ energy of \sqrts{7}{\TeV},
which amounts to an integrated luminosity of 4.7 \ifb.

A kinematic fit is used to fully reconstruct the top quark pair in each event,
where the mass values of the \tquark\ and the \antitquark\ are obtained from this fit.
Therefore, the difference between the \tquark\ and \antitquarka\ masses can be defined as:
\begin{equation}
  \ensuremath{\mDiffFit=q_{\ell} \cdot \left(\mass{b\ell\nu}{fit}-\mass{bjj}{fit}\right)},
\end{equation}
where \ensuremath{q_{\ell}} is the charge of the lepton.
As one of the \tquark s in the \ttbar\ pair decays hadronically and the other one decays leptonically,
\mass{b\ell\nu}{fit}\ is the fitted mass of the leptonic decay
while \mass{bjj}{fit}\ is the fitted mass of the hadronic decay.

The distribution of \mDiffFit\ is parametrized as the sum of two Gaussians
for the signal, and as a single Gaussian function for the background.
The signal templates are shown in \refFig{diffTemplate}.

%

\begin{figure}[h]
  \includegraphics[width=\kFigWidth\textwidth]{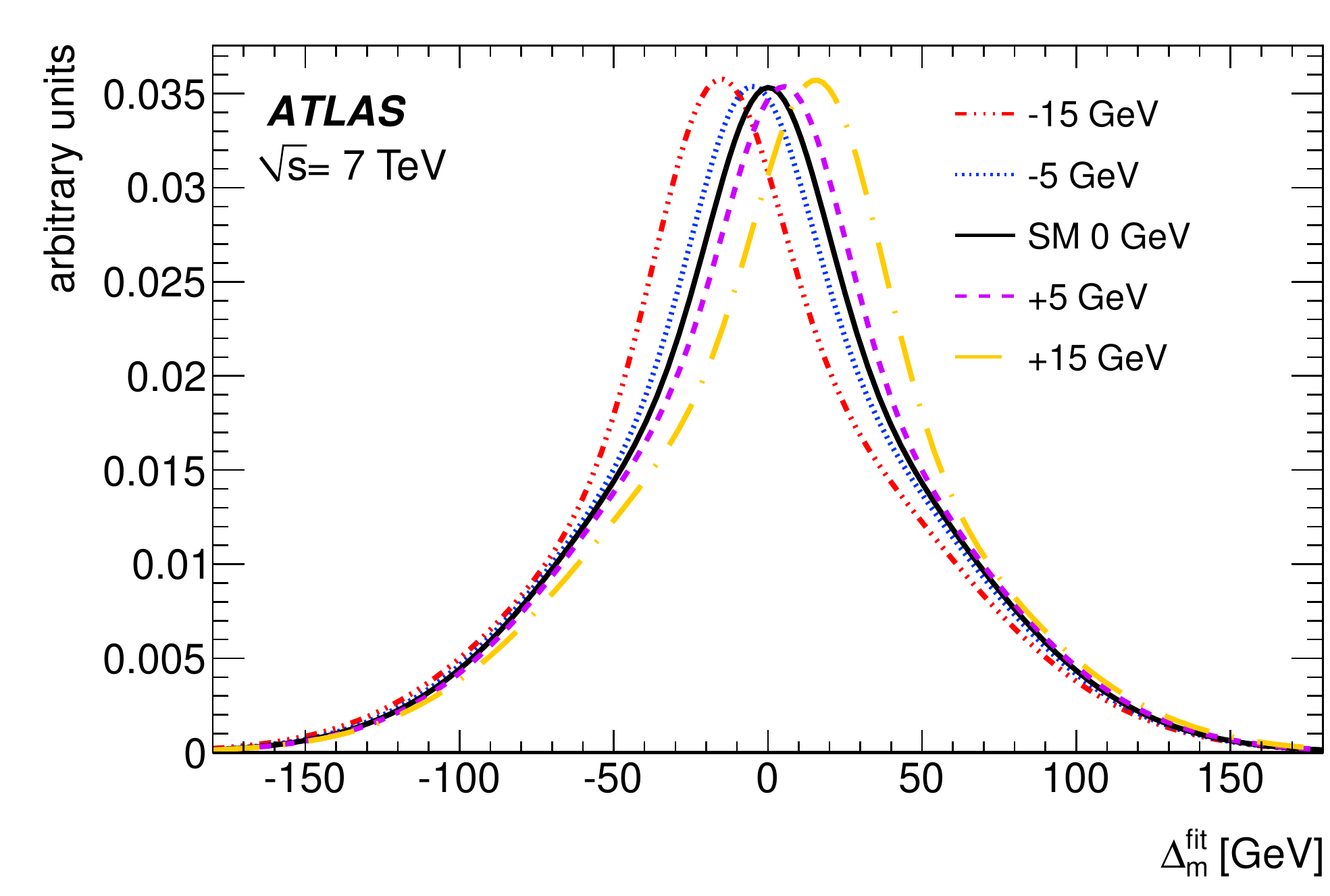}
	\caption{Signal template for the \mDiffFit\ distribution.  The distribution is modelled as the sum of two Gaussian functions.}
	\label{diffTemplate}
\end{figure}

\begin{figure}[h]
  \includegraphics[width=\kFigWidth\textwidth]{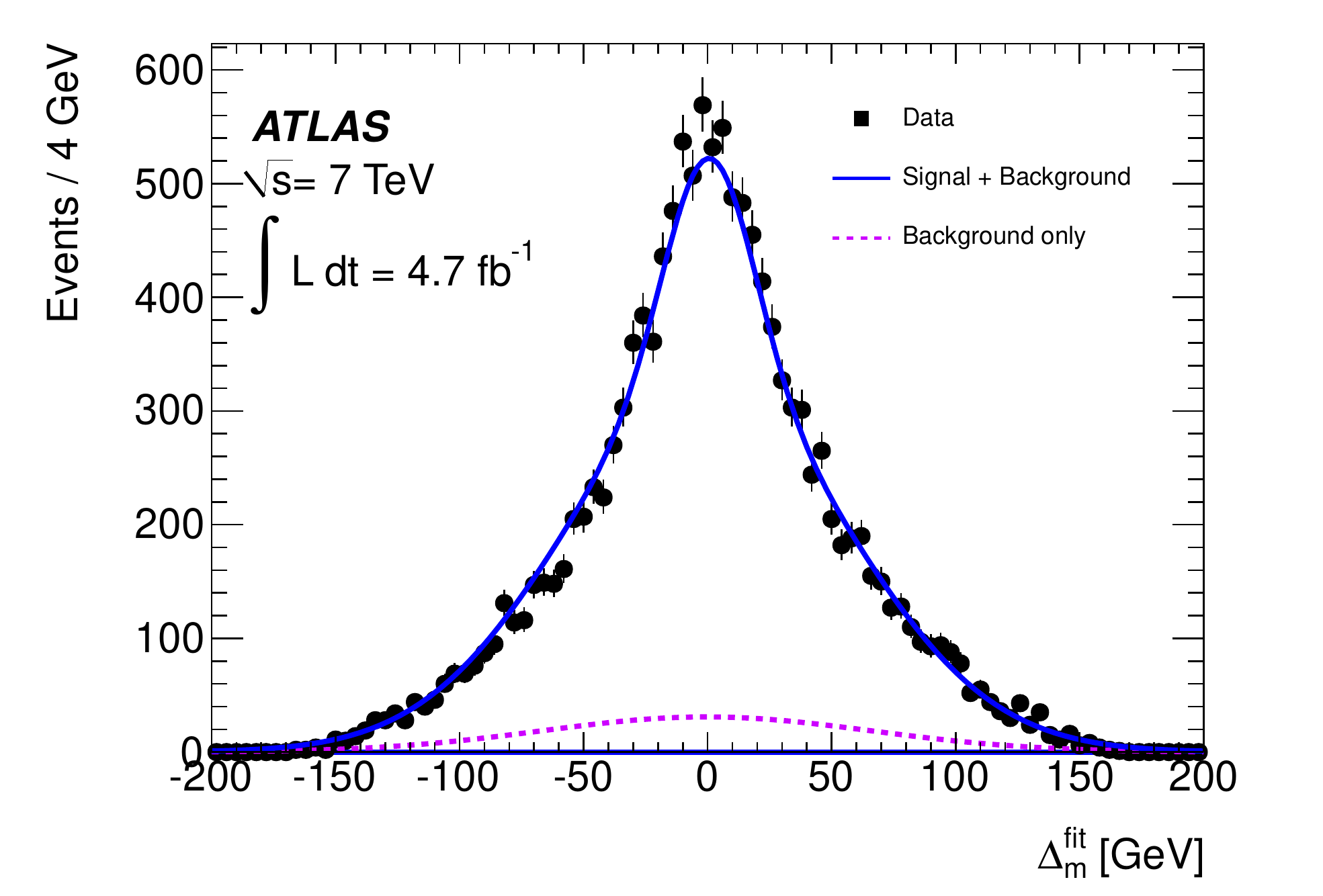}
	\caption{Reconstructed \tantit\ mass difference distribution in data, fitted with the template (signal and background).}
	\label{diffMeasurement}
\end{figure}

Applying the template fit to data (see \refFig{diffMeasurement}), the measured mass difference between the \tquark\ and the \antitquark\ is:
\begin{equation}
  \ensuremath{\Delta m_{\ttbar} = 0.67 \pm 0.61\mathrm{(stat)}\pm 0.41\mathrm{(syst)} \GeV}.
\end{equation}

The systematic uncertainty is completely dominated by the $b$-fragmentation model.
This uncertainty addresses differences in the detector response to jets originating from $b$ and $\bar{b}$ quarks.
In order to evaluate this effect, events were generated using \textsc{Powheg} and interfaced with \textsc{Pythia}.
These events were then compared with the same generated events, but the $b$-hadron decays were simulated with \textsc{EvtGen} instead.
Using this procedure, the $b$-fragmentation model uncertainty is estimated to be 0.34 \GeV.

\subsection*{Acknowledgment}
The research leading to these results has received funding from the European Research Council under the European Union's Seventh Framework Programme ERC Grant Agreement n. 617185.




\nocite{*}
\bibliographystyle{elsarticle-num}
\bibliography{biblio}

\begin{thebibliography}{1}
\expandafter\ifx\csname url\endcsname\relax
  \def\url#1{\texttt{#1}}\fi
\expandafter\ifx\csname urlprefix\endcsname\relax\def\urlprefix{URL }\fi
\expandafter\ifx\csname href\endcsname\relax
  \def\href#1#2{#2} \def\path#1{#1}\fi

\bibitem{ATLAS}
\ATLASC, {The ATLAS Experiment at the CERN Large Hadron Collider}, JINST 3~(08)
  (2008) S08003.

\bibitem{LHC}
{L. Evans and P. Bryant (editors)}, {LHC Machine}, JINST 3~(08) (2008) S08001.

\bibitem{SingleLepton}
\ATLASC, \href{http://cds.cern.ch/record/1547327}{{Measurement of the Top Quark
  Mass from $\sqrt{s}=7$ TeV ATLAS Data using a 3-dimensional Template
  Fit}}~(ATLAS-CONF-2013-046).
\newline\urlprefix\url{http://cds.cern.ch/record/1547327}

\bibitem{Dilepton}
\ATLASC, \href{http://cds.cern.ch/record/1562935}{{Measurement of the Top Quark
  Mass in Dileptonic Top Quark Pair Decays with $\sqrt{s}=7$ TeV ATLAS
  Data}}~(ATLAS-CONF-2013-077).
\newline\urlprefix\url{http://cds.cern.ch/record/1562935}

\bibitem{PoleMass}
\ATLASC, {Measurement of the $t\bar{t}$ production cross-section using $e\mu$
  events with $b$-tagged jets in $pp$ collisions at $\sqrt{s}=7$ and 8 TeV with
  the ATLAS detector}~(CERN-PH-EP-2014-124).
\newblock \href {http://arxiv.org/abs/1406.5375} {\path{arXiv:1406.5375}}.

\bibitem{MassDiff}
\ATLASC, Measurement of the mass difference between top and anti-top quarks in
  pp collisions at using the {ATLAS} detector, Phys. Lett. B 728 (2014) 363 --
  379.

\bibitem{TopMassMCVsPole}
A.~Hoang, I.~Stewart, {Top Mass Measurements from Jets and the Tevatron
  Top-Quark Mass}, Nucl. Phys. B (Proc. Suppl.) 185 (2008) 220--226.

\end{thebibliography}







\end{document}
